\documentclass[12pt]{article}
\usepackage{latexsym,amssymb,amsmath}

\begin{document}

\title
{Multi-instantons in seven dimensions}
\author
{E.K. Loginov\footnote{Research supported by RFBR Grant
04-02-16324;
E-mail: loginov@ivanovo.ac.ru}\\
Physics Department, Ivanovo State University\\
Ermaka St. 39, Ivanovo, 153025, Russia}
\date{February 19, 2005}
\maketitle

\begin{abstract}
We consider the self-dual Yang-Mills equations in seven
dimensions. Modifying the t'Hooft construction of instantons in
$d=4$, we find $N$-instanton $7d$ solutions which depend on $8N$
effective parameters and are $E_{6}$-invariant.
\end{abstract}

\section{Introduction}

The pure Yang-Mills (YM) theory defined in the four-dimensional
Euclidean space has a rich and interesting structure even at the
classical level. The discovery of regular solutions to the YM
field equations, which correspond to absolute minimum of the
action (Belavin et al.)~[1], has led to an intensive study of such
a classical theory. One hopes that a deep understanding of the
classical theory will be invaluable when one tries to quantize
such a theory.
\par
In the past few years, increased attention has been paid to gauge
field equations in space-time of dimension greater than four, with
a view to obtaining physically interesting theories via
dimensional reduction~[2]. Such equations appear in the
many-dimensional theory of supergravity, in the low-energy
effective theory of $d$-branes, and in M-theory~[3]. Using
solutions of the YM equations in $d>4$ makes possible to obtain
soliton solutions in these theories~[4]. It is known also that the
YM theory in $d$ dimensions may be reduced to the Yang-Mills-Higgs
(YMH) theory in $k<d$ dimensions~[5]. Hence, solutions of the YMH
equations in $d=4$ may be obtained from solutions of the YM
equations in $d>4$ dimensions.
\par
In Ref.~[6], the 4d self-dual Yang-Mills equations was generalized
to the higher-dimensional linear relations (CDFN equations)
\begin{equation}
c_{mnps}F^{ps}=\lambda F_{mn},
\end{equation}
where the numerical tensor $c_{mnps}$ is completely antisymmetric
and $\lambda=const$ is a non-zero eigenvalue. It is obviously that
these equations lead to the full YM equation, via the Bianchi
identity. Several self-dual solutions of (1) were found in~[7].
\par
The paper is organized as follows. Sections 2 and 3 contain
well-known facts about the Cayley-Dickson algebras and their
derivations. In Section 4 multi-instanton solutions of the
$G_2$-invariant CDFN equations are found. In Section 6 the
$E_{6}$-invariance of these solutions are proved.

\section{Cayley-Dickson algebras}

Let $A$ be an algebra with an involution $x\to\bar x$ over a field
$F$ of characteristic $\ne 2$. Given a nonzero $\alpha\in F$ we
define a multiplication on the vector space $(A,\alpha)=A\oplus A$
by
\begin{equation}
(x_1,y_1)(x_2,y_2)=(x_1x_2-\alpha \bar y_2y_1,y_2x_1+y_1\bar x_2).
\notag
\end{equation}
This makes $(A,\alpha)$ an algebra over $F$. It is clear that $A$
is isomorphically embedded into $(A,\alpha)$ and
$\text{dim}(A,\alpha)=2\text{dim}A$. Let $e=(0,1)$. Then
$e^2=-\alpha$ end $(A,\alpha)=A\oplus Ae$. Given any $z=x+ye$ in
$(A,\alpha)$ we suppose $\bar z=\bar x-ye$. Then the mapping
$z\to\bar z$ is an involution in $(A,\alpha)$.
\par
Starting with the base field $F$ the Cayley-Dickson construction
leads to the following sequence of alternative algebras:
\par
1) $F$, the base field.
\par
2) $\mathbb C(\alpha)=(F,\alpha)$, a field if $x^2+\alpha$ is the
irreducible polynomial over $F$; otherwise, ${\mathbb
C}(\alpha)\simeq F\oplus F$.
\par
3) $\mathbb H(\alpha,\beta)=(\mathbb C(\alpha),\beta)$, a
generalized quaternion algebra. This algebra is associative but
not commutative.
\par
4) $\mathbb O(\alpha,\beta,\gamma)=(\mathbb
H(\alpha,\beta),\gamma)$, a Cayley-Dickson algebra. It is easy to
prove that this algebra is nonassociative.
\par
The algebras in 1) -- 4) are called composition. Any of them has
the non\-degenerate quadratic form (norm) $n(x)=x\bar x$, such
that $n(xy)=n(x)n(y)$. The norm $n(x)$ defines the scalar product
\begin{equation}
(x,y)=\frac12(\bar xy+\bar yx)
\end{equation}
that is invariant with respect to all automorphisms of the
composition algebra. It is known also that over the field $\mathbb
R$ of real numbers, the above construction gives 3 split algebras
(e.g., if $\alpha=\beta=\gamma=-1$) and 4 division algebras (if
$\alpha=\beta=\gamma=1$): the fields of real $\mathbb R$ and
complex $\mathbb C$ numbers, the algebras of quaternions $\mathbb
H$ and octonions $\mathbb O$, taken with the Euclidean norm
$n(x)$. Finally note that any composition algebra is alternative,
i.e. in any of them the associator
\begin{equation}
(x,y,z)=(xy)z-x(yz)
\end{equation}
is skew-symmetric over $x,y,z$. Note also that any simple
nonassociative alternative algebra is isomorphic to the
Cayley-Dickson algebra $\mathbb O(\alpha,\beta,\gamma)$.
\par
As any finite-dimensional algebra, the Cayley-Dickson algebra may
be defined by "a multiplication table" in some fixed basis. For
that we consider a real linear space $A$ equipped with a
nondenerate symmetric metric $g$ of signature $(8,0)$ or $(4,4)$.
Choose the basis $1,e_{1},\dots,e_{7}$ in $A$ such that
\begin{equation}
g=\text{diag}(1,\alpha,\beta,\alpha\beta,\gamma,\alpha\gamma,\beta\gamma,\alpha\beta\gamma),
\end{equation}
where $\alpha,\beta,\gamma=\pm1$. Define the multiplication
\begin{equation}
e_{i}e_{j}=-g_{ij}+c_{ij}{}^{k}e_{k},
\end{equation}
where the structural constants $c_{ijk}=g_{ks}c_{ij}{}^{s}$ are
completely antisymmetric and different from 0 only if
\begin{equation}
c_{123}=c_{145}=c_{167}=c_{246}=c_{275}=c_{374}=c_{365}=1.
\nonumber
\end{equation}
The multiplication (5) transform $A$ into a real linear algebra.
It can easily be checked that $A$ is isomorphic to $\mathbb
O(\alpha,\beta,\gamma)$.

\section{Derivations}

Recall that a derivation of an algebra $A$ is a linear
transformation $D$ of $A$ satisfying
\begin{eqnarray}
(xy)D=(xD)y+x(yD) \nonumber
\end{eqnarray}
for all $x,y\in A$. The derivations of Cayley-Dickson algebra may
be described in intrinsic terms. Namely, let $A$ be a
Cayley-Dickson algebra. Then for any $x,y\in A$ the mapping
\begin{eqnarray}
D_{x,y}:z\to 2[z,[x,y]]+6(z,x,y)
\end{eqnarray}
is a derivation of $A$. Therefore we have the linear mapping
$\Lambda^2\to \text{Der}A$. Since any Cayley-Dickson algebra is
simple, it follows that this mapping is surjective. In addition,
the following relations
\begin{align}
D_{x,yz}&=D_{y,xz}+D_{z,yx},\\
[D_{x,y},D_{z,t}]&=D_{(xD_{z,t}),y}+D_{x,(yD_{z,t})}
\end{align}
are true. Note also that the derivations algebra $\text{Der}A$ is
a simple exceptional Lie algebra of type $g_2$.
\par
Since the associator (3) of Cayley-Dickson algebra is
skew-symmetric over its arguments, it follows that we can define
the completely antisymmetric tensor $c_{ijkl}$ by
\begin{align}
(e_i,e_j,e_k)=2c_{ijk}{}^le_l.
\end{align}
It is easy to prove that this tensor satisfies the following
identities:
\begin{align}
c_{ijs}c_{kl}{}^{s}&=g_{ik}g_{jl}-g_{il}g_{jk}+c_{ijkl},\\
c_{ijps}c_{kl}{}^{ps}&=4(g_{ik}g_{jl}-g_{il}g_{jk})+2c_{ijkl};
\end{align}
and has the nonzero components:
\begin{eqnarray}
c_{4567}=c_{2367}=c_{2345}=c_{1357}=c_{1364}=c_{1265}=c_{1274}=1.
\nonumber
\end{eqnarray}
Further, it follows from (2) and (9) that the tensor $c_{ijkl}$ is
invariant with respect to all automorphisms of algebra $A$. Noting
that the group $\text{AutA}$ is isomorphic to the Lie group of
type $G_2$, we see that the tensor $c_{ijkl}$ is $G_2$-invariant.
Finally, rewriting the identity (7) in the form
\begin{eqnarray}
c_i{}^{jk}D_{jk}=0, \nonumber
\end{eqnarray}
where the derivation $\frac18D_{e_i,e_j}$ is denoted by the symbol
$D_{ij}$, we get the following relations
\begin{eqnarray}
c_{ij}{}^{kl}D_{kl}=-2D_{ij}.
\end{eqnarray}
\par
Since the algebra $\text{Der}A$ is a Lie algebra of type $g_2$ (or
$g'_2$ in noncompact case), it follows that it may be considered
as a subalgebra of the Lie algebra $so(m,n)$ of type $so(7)$ or
$so(3,4)$. Hence there exists the projector $c^+_{ijkl}$ of one
onto the subspace $\text{Der}A$. Usually this projector is chosen
in the form~(See~[7]):
\begin{eqnarray}
c^+_{ijkl}=\frac16\left(2g_{ik}g_{lj}-2g_{il}g_{jk}-c_{ijkl}\right).
\end{eqnarray}
In addition, it is easily shown that the derivations
\begin{eqnarray}
D_{ij}=\frac32c^+_{ij}{}^{kl}E_{kl},
\end{eqnarray}
where $E_{kl}$ are generators of the Lie algebra $so(m,n)$
satisfying the switching relations
\begin{eqnarray}
[E_{ij},E_{kl}]=g_{k[i}E_{j]l}-g_{l[i}E_{j]k}. \nonumber
\end{eqnarray}
Besides, it follows from (11) that
\begin{eqnarray}
c_{ij}{}^{ps}c^+_{klps}=-2c^+_{ijkl}.
\end{eqnarray}
Comparing (14) and (15), we again obtain the identity (12).

\section{Solutions}

Recall that the self-dual equations has been successfully tackled
by the twis\-tor techniques, and in the case of finite action
solutions by the algebraic ADHM construction~[9]. A generalization
of the ADHM construction for the equations (1) which break
$SO(4n)$ up to $Sp(1)\times Sp(n)/Z_{2}$ was found in~[10].
However in dimensions $7$ and $8$ there exists an exceptional
$G_{2}$-covariant (respectively $Spin(7)$-covariant) duality which
is connected with the octonionic algebra. Therefore the search of
generalized ADHM construction in $d=7$ and $8$ appears very
attractive.
\par
Such attempt was done in the recent paper~[11]. In one the
generalized ADHM construction in $d=8$ was built with the help of
the algebra $L(\mathbb O)$ of left multiplications of octonionic
algebra $\mathbb O$. Unfortunately, calculating the field strength
in Section 5 and proving its self-duality the authors incorrect
use the equality $L(L(xy)z)=L(xyz)$, where $L(xyz)=x(yz)$ and
$x,y,z\in\mathbb O$. By associativity of the octonionic algebra it
would not be done.
\par
Nevertheless, it is easy to get multi-instanton solutions (but not
a generalized ADHM construction) of CDFN equations in seven
dimensions. We choose the ansatz $A_m$ in the form:
\begin{eqnarray}
A_m=\frac{\lambda^{\dag} y{}^{i}}{1+y^{\dag}y} D_{mi},
\end{eqnarray}
where $y$ is a column vector with the elements $y_1,\dots,y_N$ of
Cayley-Dickson algebra such that
$$
\begin{aligned}
y^{\dag}&=(y^k_1,\dots,y^k_N)\bar e_k,&\qquad y^k_I&\in\mathbb R,\\
\lambda^{\dag}&=(\lambda_1,\dots,\lambda_N),&\qquad \lambda_I&\in\mathbb  R^+,\\
y^k_I&=(b^k_{IJ}+\delta_{IJ}x^k)\lambda_J,&\qquad
b^k_{IJ}&=b^k_{JI}.
\end{aligned}
$$
Using the identities (8)--(10), we get the field strength
\begin{eqnarray}
F_{mn}=-\frac{\lambda^{\dag}\{(2+2y^{\dag}y-y^{i}y_{i}^{\dag})
D_{mn} +3c^+_{mn}{}^{is}
D_{sj}y^jy_i^{\dag}\}\lambda}{(1+y^{\dag}y)^2}, \nonumber
\end{eqnarray}
where the tensor $c^+_{ijkl}$ is defined dy the equality (13). Now
it follows from (12) and (15) that the field strength $F_{mn}$
satisfies the CDFN equations (1) as for Euclidean as for
pseudoeuclidean metric of the form (4).
\par
This construction of multi-instanton solutions of the CDFN
equations may be easy to extend in eight dimensions. It is
sufficient to take the projector $f^+_{ijkl}$ of the algebra Lie
of type $so(8)$ or $so(4,4)$ onto the subalgebra $so(7)$ or
$so(3,4)$ respectively in place $c^+_{ijkl}$, to define the
elements $D'_{ij}$ of the form (14), and to prove an analog of the
identity (15)~(See~[7]). Then choosing the ansatz $A'_m$ in the
form:
\begin{eqnarray}
A'_m=\frac{\lambda^{\dag} y{}^{i}}{1+y^{\dag}y}D'_{mi},
\end{eqnarray}
where the indexes $m,i\in\{0,\dots,7\}$, we can obtain the
following expression for the field strength:
\begin{eqnarray}
F'_{mn}=-\frac13\frac{\lambda^{\dag}\{(6+6y^{\dag}y-3y^{i}y_{i}^{\dag})
D'_{mn} +8f^+_{mn}{}^{is}
D'_{sj}y^jy_i^{\dag}\}\lambda}{(1+y^{\dag}y)^2}. \nonumber
\end{eqnarray}
Obviously, the $N$-instanton solutions (16) and (17) depend on
$8N$ and $9N$ effective parameters respectively, and are a
generalization of the t'Hooft solution in $d=4$~(see e.g.~[12]).

\section{$E_{6}$-invariance}

Let $A$ be a real Cayley-Dickson algebra with the involution
$x\to\bar x$, and let $A_3$ be the algebra of all $3\times 3$
matrix with elements of $A$. Consider the set
\begin{eqnarray}
J=\{(x_{ij})\in A_3\mid(\bar x_{ij})=(x_{ji})\}. \nonumber
\end{eqnarray}
The set $J$ is a commutative nonassociative algebra with the
respect to the product
\begin{eqnarray}
x\circ y=\frac12(xy+yx). \nonumber
\end{eqnarray}
The algebra $J$ satisfies the identity
\begin{eqnarray}
(x^2y)x=x^2(yx) \nonumber
\end{eqnarray}
and is said to be an exceptional Jordan algebra.
\par
Denote $3\times 3$ matrix $(x_{ij})$ with the unique nonzero
element $x_{ij}=1$ by the symbol $\varepsilon_{ij}$ and choose in
$J$ the basis:
\begin{eqnarray}
\aligned E_1&=\varepsilon_{11},\qquad
X_1(e_i)=e_i\varepsilon_{23}+\bar e_{i}\varepsilon_{32},\\
E_2&=\varepsilon_{22},\qquad
X_2(e_j)=e_j\varepsilon_{31}+\bar e_{j}\varepsilon_{13},\\
E_3&=\varepsilon_{33},\qquad X_3(e_k)=e_k\varepsilon_{12}+\bar
e_{k}\varepsilon_{21},
\endaligned
\end{eqnarray}
where $e_0=1,e_{1},\dots,e_{7}$ is the standard basis of $A$. It
can easily be checked that
\begin{eqnarray}
E_{\alpha}\circ X_{\beta}(e_i)= \left\{ \aligned
0,\quad&\text{if}\quad\alpha=\beta,\\
\frac12X_{\beta}(e_i),\quad&\text{if}\quad\alpha\ne\beta,
\endaligned
\right.
\end{eqnarray}
\begin{eqnarray}
X_{\alpha}(e_i)\circ X_{\beta}(e_j)= \left\{ \aligned
\delta_{ij}(E-E_{\alpha}),\quad&\text{if}\quad\alpha=\beta,\\
\frac12X_{\gamma}(\bar e_j\bar
e_i),\quad&\text{if}\quad\alpha\ne\beta,
\endaligned
\right.
\end{eqnarray}
where $E$ is the identity $3\times 3$ matrix, and
$(\alpha\beta\gamma)=(123),(231),(312)$.
\par
It is well known~(see e.g.~[8]) that the derivations algebra
$\text{Der}J$ is a simple exceptional Lie algebra of the type $f_4$.
Since there is an isomorphic enclosure of the algebra $g_{2}$ into
$f_{4}$, we can consider (16) as a field that takes its values in
$\text{Der}J$. To prove the $F_{4}$-invariance of
these solutions, we find the trace of the matrix
\begin{eqnarray}
X_{\beta}=\{(X_{\alpha}(e_i),X_{\beta}(e_j),X_{\alpha}(e_k))
-\frac12(X_{\alpha}(e_i),X_{\alpha}(e_j),X_{\alpha}(e_k))\}\circ
X_{\beta}(e_l),
\end{eqnarray}
where $i,j,k\ne 0$, and we do not sum on the recurring indexes. Using
(9) and (19)--(20), we prove that
\begin{eqnarray}
X_{\beta}=\frac12c_{ijkl}(E-E_{\beta}), \nonumber
\end{eqnarray}
and hence
\begin{eqnarray}
\text{tr}X_{\beta}=c_{ijkl}. \nonumber
\end{eqnarray}
Since a trace of matrix in $J$ is invariant with respect to all
automorphisms of $J$, we prove the $F_4$-invariance of solutions
of the corresponding CDFN equations.
\par
Moreover, it can be proved that the tensor $c_{ijkl}$ is
$E_6$-invariant. Indeed, the group $E_6$ is a group of linear
transformations of the space $J$ that preserve the norm
\begin{align}
n(X)&=x_{11}x_{22}x_{33}+(x_{12}x_{23})x_{31}+x_{13}(x_{32}x_{21})
\nonumber\\
&-x_{11}x_{23}x_{32}-x_{22}x_{31}x_{13}-x_{33}x_{12}x_{21},
\nonumber
\end{align}
where $X=(x_{ij})\in J$. Choose an element $X$ in the form
\begin{eqnarray}
X=X_1+X_2-X_3+E_1+E_2, \nonumber
\end{eqnarray}
where matrixes $E_{\alpha}$ and $X_{\beta}$ are defined by the
relations (18) and (21) respectively. Then it follows easily that
the norm
\begin{eqnarray}
n(X)=c_{ijkp}. \nonumber
\end{eqnarray}
Since the group $F_{4}$ can be isomorphically enclosed into the
group $E_{6}$, we prove the $E_{6}$-invariance of the found
solutions.

\end{document}